\documentclass[12pt,preprint]{aastex}
\usepackage{graphicx}

\begin{document}

\title{Substructure in the Coma Cluster: Giants vs Dwarfs}

\author{Scott A. Edwards, Matthew Colless\altaffilmark{1}}
\affil{Research School of Astronomy and Astrophysics, The Australian
National University, Weston Creek, ACT 2611, Australia}
\altaffiltext{1}{Correspondence to: colless@mso.anu.edu.au}
\author{Terry J. Bridges}
\affil{Anglo-Australian Observatory, PO Box 296, Epping, NSW 1710, Australia}
\author{Dave Carter}
\affil{Liverpool John Moores University, Astrophysics Research Institute,
Twelve Quays House, Egerton Wharf, Birkenhead, Wirral, CH41 1LD, UK}
\author{Bahram Mobasher\altaffilmark{2}}
\affil{Space Telescope Science Institute, 3700 San Martin Drive, 
Baltimore, MD 21218, USA}
\altaffiltext{2}{Also affiliated with the Space Sciences Department of the 
European Space Agency} 
\author{Bianca M. Poggianti}
\affil{Osservatorio Astronomico di Padova, vicolo dell'Osservatorio 5,
35122 Padova, Italy}

\begin{abstract}
The processes that form and shape galaxy clusters, such as infall,
mergers and dynamical relaxation, tend to generate distinguishable
differences between the distributions of a cluster's giant and dwarf
galaxies. Thus the dynamics of dwarf galaxies in a cluster can provide
valuable insights into its dynamical history. With this in mind, we look
for differences between the spatial and velocity distributions of giant
(b\,$<$\,18) and dwarf (b\,$>$\,18) galaxies in the Coma cluster. Our
redshift sample contains new measurements from the 2dF and WYFFOS
spectrographs, making it more complete at faint magnitudes than any
previously studied sample of Coma galaxies. It includes 745 cluster
members -- 452 giants and 293 dwarfs. We find that the line-of-sight
velocity distribution of the giants is significantly non-Gaussian, but
not that for the dwarfs. A battery of statistical tests of both the
spatial and localised velocity distributions of the galaxies in our
sample finds no strong evidence for differences between the giant and
dwarf populations. These results rule out the cluster as a whole having
moved significantly towards equipartition, and they are consistent with
the cluster having formed via mergers between dynamically-relaxed
subclusters.
\end{abstract}

\section{Introduction}

Galaxy clusters provide us with valuable insights into the formation of
large-scale structure in the Universe, as well as being laboratories for
studying galaxy formation and evolution. The Coma cluster is the most
heavily studied of all galaxy clusters (see \citet{Biv98} for an
overview). Several lines of evidence, mostly based around the presence
of substructure in the cluster, suggest that Coma has recently undergone
mergers with smaller clusters. The first evidence for substructure was
based on the spatial positions of galaxies on the sky \citep{Fit87,
Mel88, Esc92, Mer94}; more detailed evidence came from maps of the X-ray
emitting gas in the cluster obtained by ROSAT \citep{Bri92, Whi93,
Vik94}. Further evidence for a relatively recent merger history in Coma
also comes from its anomalously low early-type dwarf-to-giant ratio
\citep{Sek96}.

A deeper understanding of a cluster's mass distribution and dynamical
history can come from the line-of-sight velocities of the cluster
members, obtained from redshift measurements. \citet[CD96]{Col96} made
the first identification of statistically significant substructure in
Coma's velocity distribution (an earlier study by \citet{Dre88} did not
find evidence of significant velocity substructure). CD96 were also able
to develop a likely merger history for the cluster, involving three
subclusters centred around each of the major cD galaxies: NGC 4874, NGC
4889, and NGC 4839 (strictly, NBC 4889 is a D galaxy since it lacks the
extended halo characterising cD galaxies, but we will refer to all three
central dominant galaxies as cDs for convenience). \citet[B96]{Biv96}
carried out a similar analysis, although their interpretation of the
results was slightly different. Notably, they argued that subclustering
is only evident for the brighter (b\,$<$\,17) galaxies, and that the
distribution of fainter dwarf galaxies is much smoother and comprises
the ``main body'' of the cluster.
  
These studies of the kinematics of Coma have suffered from a lack of
statistically large numbers of dwarf galaxy redshifts in their samples.
This is important, since the kinematics of the dwarfs may provide some
important clues towards the dynamics and formation of the cluster. For
example, clusters acquire full dynamical relaxation in two stages. The
first is through violent relaxation, which leaves all galaxies with the
same velocity dispersion regardless of mass. The second is through
dynamical friction, which acts on longer timescales than violent
relaxation, and results in equipartition of energy. This leaves the
massive giants with a lower velocity dispersion than the dwarfs, so the
evidence of equipartition can be detected by directly comparing the
dwarf and giant distributions. The kinematics of the dwarfs might also
reveal something about their origins. For example, some subset of the
dwarfs may be preferentially accreted at late times from the field
\citep{Don95, Bra00, Gal01}, so that rather than being in virial
equilibrium they are falling into or through the cluster, either on
their first approach \citep[CD96]{Cal93} or having already passed once
through the cluster core \citep{Bur94, Vol01}. Alternatively, some of
dwarfs may in fact belong to satellite populations bound to the cD
galaxies in the clusters. In all these cases the kinematics of the
dwarfs provide a tell-tale signature. Recent studies that have searched
for these signatures in the dwarf populations of clusters other than
Coma include \citet{Ste97} on Centaurus, \citet{Dri01} on Fornax, and
\citet{Con01} and \citet{Gal01} on Virgo.

We have assembled a redshift sample that is more complete at faint
magnitudes than any used previously. This was made possible by utilising
the ability of the 2dF (2-degree Field -- see \citealt{Lew01}) and
WYFFOS \citep{Bri98} multi-fibre spectrographs to measure many spectra
at once over a wide field of view. This sample is described in section
2. In section 3 we describe how we used this sample to compare the
velocity and spatial distributions of giant and dwarf galaxies. We find
that the line-of-sight velocity distribution for the giants is
non-Gaussian with $>$99\% significance, while the dwarf distribution is
close to being a Gaussian. However, a battery of statistical tests
failed to detect any other significant evidence for differences between
the distributions of giants and dwarfs. In section 4, we show that the
lack of detectable differences in the localised velocity distribution
rules out the possibility of the entire cluster having proceeded
significantly towards equipartition, which is consistent with the
estimated timescales for dynamic friction to act in the cluster as a
whole. However, we find that the number of available redshifts for dwarf
galaxies in Coma is still not enough to distinguish more subtle
differences between the giant and dwarf distributions that may have
developed in the past due to mergers with smaller clusters.

\section{Data}

The positions, b magnitudes, and b$-$r colours of the galaxies in our
sample were taken from the GMP catalogue \citep{God83}. Our redshifts,
1174 in total, comprise the set used by CD96 (a mixture of values taken
from the literature and values measured with the Hydra spectrograph)
plus 272 more recently measured redshifts from the 2dF spectrograph at
the Anglo Australian Telescope. A further 123 redshifts were taken from
the sample measured using the WYFFOS spectrograph on the William
Herschel Telescope \citep{Kom01, Mob01, Pog01}. 76 of the galaxies
observed with WYFFOS were also observed with 2dF, and the corresponding
redshifts were averaged; the other 47 were not covered by the 2dF data.
Due to these extra redshifts from 2dF and WYFFOS our sample is more
complete at faint magnitudes than the sample used by CD96.

Because of the low resolution of available images of the dwarf galaxies
in Coma, it was not possible to identify them morphologically. Instead,
we distinguish dwarfs from giants using a magnitude cutoff. The choice
of a cutoff is arbitrary, as there is no obvious break in the properties
of the galaxies in our sample as a function of magnitude. We chose an
absolute dividing magnitude of $-$17, which is consistent with values
used in earlier papers --- e.g.\ \citet{Pog01}. At the distance of the
Coma cluster, this corresponds to an apparent magnitude of b\,=\,18
(assuming H$_0~{\sim}~70$~km~s$^{-1}$~Mpc$^{-1}$). In order to determine
the completeness of the giant and dwarf samples, we compare the number
of galaxies in the redshift samples with the total number in the GMP
catalogue (complete to b\,=\,20), excluding those that fall outside the
red edge of the cluster colour-magnitude relation, which is described in
section 3.3 and Fig. 5. Galaxies bluer than the bottom edge of the
colour-magnitude band were left in the sample, since there is a
reasonable chance that unusually blue galaxies may be cluster members
that have only recently fallen into the cluster. As such, the
completeness percentages quoted below are lower limits on the true
completeness. The redshift data for the `giant' sample (b\,$<$\,18) is
80\% complete over the whole cluster, and 100\% complete within 20$'$ of
the the cluster centre (which is approximately at the position of NGC
4874). The set of redshifts for galaxies with b\,$>$\,18 is only 34\%
complete over the whole cluster but 55\% complete in the central 20$'$.
To check for any spatial bias in these dwarf redshifts, we first noted
that the limited sample 18\,$<$\,b\,$<$\,19 is about 60\% complete over
the whole cluster and 85\% complete in the central region. Thus any bias
present for this limited sample would be small, especially in the
central region. We then compared this limited sample with the spatial
distribution of galaxies fainter than b\,=\,19. We found that redshift
measurements for the b\,$>$\,19 galaxies were restricted mainly to the
central regions near the cD galaxies. This is an artifact of low
completeness away from the centre of the cluster -- the b $>$ 19
subsample is 39\% complete in the central 20$'$ but only 22\% complete
over the central 50$'$. Thus there is some bias towards the cluster
centre in the dwarf sample. However, all of the statistical tests
discussed in this paper were duplicated using the
upper-magnitude-limited 18\,$<$\,b\,$<$\,19 dwarf sample, and none of
the results differed significantly from those obtained using the
b\,$>$\,18 sample (with no upper magnitude limit). We concluded that the
spatial bias in the dwarf redshift sample does not significantly affect
our results. For the rest of the paper, the term `dwarf sample' is
assumed to refer to galaxies with b $>$ 18, with no enforced upper
magnitude limit.

Galaxies with redshifts in the range
(4000\,$<$\,\textit{cz}\,$<$\,10000)~km~s$^{-1}$ were taken to be
members of the cluster -- others were assumed to be field galaxies and
were excluded. Over the whole cluster, the resulting giant sample
contains 452 galaxies and the dwarf sample contains 293 galaxies. In the
following we focus mostly on the region within 20$'$ of the cluster
centre, as this was where our redshift sample was most complete. It
should be pointed out that the most significant substructure found by
CD96 was about 40$'$ from the cluster centre -- the subcluster
surrounding NGC 4839, which appears to be in the process of colliding
with the rest of the cluster. We devote some attention to this region in
the section on spatial substructure below.

\section{Results}

\subsection{Line-of-sight velocity distributions}

CD96 found that the line-of-sight velocity distribution for galaxies in
Coma was non-Gaussian with a $>$99\% probability. However, their sample
contained few dwarfs -- it was roughly the same set of galaxies as our
sample of giants. To investigate whether this non-Gaussianity is also
apparent for the dwarfs, we plot histograms of the velocity
distributions of the giants and the dwarfs separately in Fig. 1. At a
glance the giant distribution seems to be non-Gaussian, and a Lilliefors
test (a Kolmogorov-Smirnov [K-S] test which takes into account the fact
that the mean and standard deviation are estimated from the data)
confirms this impression, rejecting a Gaussian fit with $>$99\%
probability. The dwarf distribution, on the other hand, is only rejected
as a Gaussian with less than 75\% confidence. Thus, there is strong
evidence from the velocities alone for substructure in the giant
distribution, but not for the dwarfs.

Despite this hint that the giant and dwarf distributions are different,
a direct comparison between the two using a two-sample K-S test gives
only a 45\% chance that this is actually the case. It may be that there
simply are not enough dwarf redshifts to be able to tell whether the
dwarf velocities are drawn from the same distribution as the giants or
whether they really do in fact follow a Gaussian. To check this, a
sample of 293 giants (as many as the number of dwarfs) was selected at
random from the giant distribution and tested for Gaussianity. This was
repeated 1000 times. We found that 3\% of the random samples generated
in this way had a $<$75\% chance of being non-Gaussian. Thus there is a
small but not negligible possibility that the dwarfs could be drawn from
the same distribution as the giants, and still not show any detectable
signs of non-Gaussianity. We had to increase the number of sampled
giants to about 330 for there to be a less than 1\% chance of a sample
having a $<$75\% of being non-Gaussian. Thus we would have to increase
the number of dwarf redshifts in our sample before we can tell with any
certainty if the line-of-sight velocity distributions really are
different.

Given the arbitrary nature of our choice for a dividing magnitude
between giants and dwarfs, it is worth investigating whether changing
the dividing magnitude affects these results. If b\,=\,17 is used instead
of b\,=\,18, the number of giants and dwarfs becomes 306 and 439
respectively -- note that there are now more dwarfs than giants. Despite
this, the results of the Lilliefors tests and the two-sample K-S test
are exactly the same as before -- the giants are significantly
non-Gaussian but not the dwarfs, while a direct comparison finds only a
45\% chance that the two distributions differ. We take this as an
indication that our results are not sensitive to small changes in the
choice of dividing magnitude. This also increases our confidence that
the non-Gaussianity in the line-of-sight velocity distribution really is
restricted to the giants, since in this case we can't put the difference
down to having less dwarf redshifts than giant redshifts. Finally, since
B96 used a cutoff of b\,=\,17 rather than 18 to divide the giants and
dwarfs, this result tells us that we can meaningfully compare our
results to theirs.

The velocity dispersion for the giants is $\sigma_{Gi}$ =
(979~$\pm$~30)~km~s$^{-1}$, and for the dwarfs $\sigma_{Dw}$ =
(1096~$\pm$~45) km s$^{-1}$. The dwarf dispersion is larger than that of
the giants at the 2.2$\sigma$ level. This is qualitatively consistent
with the effects of equipartition, which involves the transfer of
kinetic energy from massive galaxies to lighter ones. However, complete
equipartition would generate a far bigger difference between the giant
and dwarf dispersions. To demonstrate this, we performed a Monte-Carlo
simulation in which each galaxy was assigned a velocity drawn from a
Gaussian with a dispersion proportional to $1/\sqrt{M}$ (where $M$ is
the mass of the galaxy in question), as expected for a cluster in
equipartition. The mass of each galaxy is assumed to be proportional to
its luminosity, and the mass-to-light ratio is assumed to be the same
for all galaxies. After assigning each galaxy a velocity, all of the
velocities were scaled so that the dispersion of the total sample
(giants plus dwarfs) was the same as that for the real data, about 1000
km s$^{-1}$. The resulting dispersions for the giants and for the dwarfs
(averaged over five simulations) were $\sigma_{Gi}$ =
(482~$\pm$~16)~km~s$^{-1}$ and $\sigma_{Dw}$ = (1523~$\pm$~60), a much
larger difference than is seen in the real data. We can conclude from
this that the cluster as a whole can only have marginally begun to
proceed towards equipartition.

There is some evidence that many dwarf ellipticals in clusters arise
from the infall of late-type galaxies \citep{Gal01}. The infall scenario
predicts that the dwarf velocity distribution should be broader than
that of the giants. This is because an infalling population is not
virialised; the kinetic energy is roughly equal to the potential energy,
rather than half the potential energy, so the velocities of infalling
galaxies should be on average a factor of $\sqrt{2}$ larger than those
of a virialised population. As noted above, the dwarf line-of-sight
velocity distribution is indeed slightly broader than the giant
distribution. However, as well as being wider, we would expect the
distribution of an infalling population to be non-Gaussian (because it
is not dynamically relaxed), and we see no evidence of that here. The
fact that the dwarf distribution is close to Gaussian suggests that even
if the dwarfs did form from infall in the past, they have since
undergone some form of dynamical relaxation which has erased the
kinematic signature of the infall. An alternative explanation is
possible if infall occurs in bursts, rather than continuously: a
superposition of many `infall burst' distributions of various
dispersions and mean velocities could conceivably (although not
necessarily) look Gaussian.

To investigate the cause of the non-Gaussianity of the giants' velocity
distribution, the cumulative frequency of giants as a function of
\textit{cz} was plotted along with a cumulative Gaussian with \ the same
mean and dispersion (Fig. 2). We can see that the largest deviation from
a Gaussian distribution occurs at \textit{cz} $\sim$ 6500~km~s$^{-1}$.
This corresponds to the position of the largest gap in the velocity
histogram for the giants (Fig. 1). The gap is not evident in the dwarf
distribution. Remarkably, the central velocity of this gap is close to
the velocity of NGC 4889 (\textit{cz} = 6508~km~s$^{-1}$). Also, a
smaller gap around \textit{cz} = 7400~km~s$^{-1}$ seems to correspond to
the velocity of NGC 4839. This begs the question of whether or not these
gaps are directly caused by the cD galaxies, via mergers or other less
obvious dynamical effects. We can gain some insight into this question
by looking at the spatial positions of galaxies at different redshifts,
as described in the next section.

\subsection{Localised Velocity Structure}

Fig. 3 is a plot of the smoothed galaxy distribution as a function of
projected distance D along the NE-SW axis in the sky and of the
line-of-sight velocity \textit{cz}, for the giants (a) and for the
dwarfs (b). D is defined in terms of $\Delta$RA and $\Delta$Dec by D =
($\Delta$RA+$\Delta$Dec)/$\sqrt{2}$, with NE being positive and D = 0 at
the cluster centre. The NE-SW axis is a natural choice because it is the
axis on which the three cD galaxies are the most separated, and it seems
to correspond roughly to the direction along which the NGC 4839
subcluster is approaching the main body of Coma (CD96). The smoothed
density plot in Fig. 3 was generated by convolving the actual galaxy
distribution with a 2D Gaussian kernel. The smoothing algorithm was
adaptive, meaning that the width of the Gaussian was varied depending on
the local density of galaxies. The base values for the dispersions of
the Gaussian kernel were ${\sigma}_D$ = 5$'$ and
${\sigma}_{\textit{cz}}$ = 200~km~s$^{-1}$, and these were scaled for
each gridpoint by a factor proportional to $1/\sqrt{den}$, where $den$
is the density of galaxies at that gridpoint. Thus the kernel narrows in
high-density regions and broadens in low-density regions.

Recall that from the velocity distribution (Figs. 1 and 2) we expect to
see gaps in the density of giant galaxies around \textit{cz} $\sim$ 6500
and 7400~km~s$^{-1}$. There are indeed bands of noticeably reduced
density at these redshifts, but they are not strongly localised. The
only identifiable holes are at D =~$\pm$~20$'$. If the velocity gaps
were really caused by the cD galaxies, we would expect to see some
correlation between the spatial positions of these galaxies and the
gaps, but this does not appear to be the case.

When comparing these two plots by eye there seem to be obvious
differences between them, such as the peak in the density of giants at
(D~$\sim$~0$'$, \textit{cz} $\sim$ 8000~km~s$^{-1}$) which doesn't occur
in the dwarf distribution. However, these differences may well just be
statistical fluctuations. We used three different tests to calculate the
significance of any differences between the giant and dwarf
distributions: a 2D ${\chi}^2$ test, a 2D Kolmogorov-Smirnov (K-S) test,
and a Monte-Carlo resampling test. Details concerning the first two can
be found in Press et~al.\ (1992). The Monte-Carlo test works by running
through all of the galaxies in the sample and reassigning each to either
the dwarf group or the giant group at random, where the probability of
being a dwarf is just the ratio of the number of dwarfs to total number
of galaxies in the real sample. A large number of these reassigned
samples ($\sim$ 1000) are generated, and for each the binned statistic
\textit{Diff} is calculated. \textit{Diff} is defined by
\begin{equation}
\mathit{Diff} = \sum_{i}\left(\frac{D_i}{D_i + G_i} - P\right)^2
\end{equation}
where $D_i$ and $G_i$ are the numbers of dwarfs and giants in bin $i$
and $P = D_{tot}/(D_{tot}+G_{tot})$ is the probability that any given
galaxy is a dwarf. We then compare the value of \textit{Diff} for the
real sample with the range calculated for the Monte-Carlo reassigned
samples to obtain an estimate of the likelihood that the dwarfs and
giants are from different distributions. Note that the \textit{Diff}
statistic is very similar to the ${\chi}^2$ statistic; the difference is
that for \textit{Diff} each term in the sum is not weighted by the total
number of galaxies in that bin, so the result is less sensitive to small
differences in highly populated bins than would be the case for
${\chi}^2$.

Each test was applied firstly to all the galaxies with $-50'<D<50'$ and
$5000<cz<9000$~km~s$^{-1}$ (henceforth referred to as the \emph{large}
range), and then again to just the galaxies with $-20'<D<20'$ and
$6000<cz<8000$~km~s$^{-1}$ (henceforth referred to as the \emph{central}
range). When calculating the ${\chi}^2$ and Monte-Carlo statistics, any
bins with less than 5 galaxies total (giants and dwarfs) were neglected.
The number of bins was chosen to be the maximum for which the number of
neglected bins did not exceed 25\% of the total. This turned out to be a
6$\times$6--bin grid for the large range, and a 5$\times$5 grid for the
central range. The results for each test when applied to the
D-vs-\textit{cz} distribution, quoted as the probability that the giant
and dwarf distributions differ, are given in Table 1. In summary, these
tests give no better than a $\sim$85\% chance that the two distributions
are different, which is not enough to make any strong claims that
statistically significant differences exist.

Another way to investigate how velocity substructure correlates with
spatial position is to plot the galaxy densities on the sky for
restricted redshift ranges. We have done this for three different (but
slightly overlapping) ranges: $5400<cz<6600$~km~s$^{-1}$,
$6400<cz<7600$~km~s$^{-1}$ and $7400<cz<8600$~km~s$^{-1}$. The
motivation for choosing these three ranges is related to conflicting
claims made by CD96 and B96 regarding which velocity structures are
associated with which of the cD galaxies. Both papers agree that the
peak in galaxy density around (D $\sim$ 0$'$, \textit{cz} $\sim$
7000~km~s$^{-1}$) is associated with NGC 4874. However, they don't agree
about which galaxies are associated with NGC 4889. CD96 identify the
galaxies around (D $\sim$ 5$'$, \textit{cz} $\sim$ 8000~km~s$^{-1}$)
with NGC 4889, whereas B96 assume the relevant cluster to be that around
(D $\sim$ 5$'$, \textit{cz} $\sim$ 6000~km~s$^{-1}$).
    
The projected density plots for the three velocity ranges (referred to
henceforth as A, B and C respectively) for both giants and dwarfs are
shown in Fig. 4. The densities for these plots were smoothed using an
adaptive Gaussian kernel with base dispersions
${\sigma}_{\mathrm{RA}}$\,=\,${\sigma}_{\mathrm{Dec}}$\,=\,3$'$. The
densities in each of the three giant plots are set to the same scale -
that is, they are all normalised to the maximum value in region B. The
same is true independently for the three dwarf plots (in other words,
they have the same shading scale as each other but different to the
giant plots). $\Delta$RA and $\Delta$Dec are restricted to the range
${\pm}20'$ to focus attention to the area around NGC 4874 and NGC 4889.
In velocity range A, centred around \textit{cz} = 6000~km~s$^{-1}$, both
the giants and the dwarfs exhibit density maxima located roughly at the
positions of NGC 4874 and 4889, although offset by up to 10$'$. In
velocity range B, the density maxima near NGC 4889 have vanished, but a
peak near NGC 4874 is obvious for both giants and dwarfs (although the
peak in dwarf density is slightly less well defined than for the
giants). Finally, in velocity range C we still see for the giants a peak
halfway between NGC 4874 and NGC 4889, whereas for the dwarfs there is
no major peak near either of the two supergiant galaxies. Based on these
figures, it is hard to make a clear identification of any one clump in
velocity space with NGC 4889. The lack of dwarfs near NGC 4889 in region
C seems to point to the galaxies with velocities around
\textit{cz}~$\sim$~6000~km~s$^{-1}$ (region A) being associated with NGC
4889, consistent with the assumption made by B96, but this is by no
means conclusive.

B96 also argue that significant subclustering is only evident for the
brighter (giant) galaxies, and not the fainter (dwarf) galaxies. To test
this, we used the ${\chi}^2$, K-S and Monte-Carlo tests to check for
statistically significant differences between the spatial distributions
of giants and dwarfs in each of the velocity ranges plotted in Fig. 4.
For the ${\chi}^2$ and Monte-Carlo tests, we used a 4$\times$4 grid of
bins, and we neglected any bins containing less than 3 galaxies in
total. The results (quoted as the percentage chance of the two
distributions being different) are given in Table 2. These results
clearly show no evidence of significant differences between the spatial
distribution of giants and dwarfs in any of the three redshift ranges.
In fact, the distributions of giants and dwarfs seem to be unusually
\emph{similar}, compared to what we would expect if each galaxy was
assigned to be a giant or a dwarf at random, especially in range B.
Based on these results, then, there is no evidence that subclustering is
restricted to just the giant galaxies. This is not to say that such a
difference does not exist -- merely that, if it does, our statistical
tests cannot detect it with the number of redshifts available to us. The
issue of what differences our tests are in principle capable of
detecting is discussed further in section 4.

\subsection{Spatial Substructure}

The final test we carried out was to compare the spatial distribution of
giant and dwarf galaxies, ignoring redshift entirely. The giant and
dwarf samples were expanded to include all the galaxies in Coma without
determined redshifts. Since redshift can't be used to determine whether
these galaxies are part of the cluster or not, we used the
colour-magnitude relationship instead. If the b-r colour is plotted
against magnitude b for galaxies in the cluster, the points generally
fall on a distinct band, as can be seen in Fig. 5a. Galaxies that did
not have determined redshifts, and lay outside this band on the
colour-magnitude plot (Fig. 5b), were assumed to be field galaxies and
were not included in the sample. The resulting expanded giant sample has
529 galaxies, and the expanded dwarf sample has 1041 galaxies.

The smoothed density plots of the spatial distribution of the two
expanded samples are shown in Fig. 6. The results look something like
the picture of Coma put forward by B96: the distribution of giants shows
two distinct peaks around the two central supergiants, NGC 4889 and NGC
4874, while the dwarf distribution has a peak between the two
supergiants at a position corresponding to a secondary peak in the ROSAT
X-ray map. (There is also another large peak about 10$'$ W and 5$'$ S of
NGC 4874, which doesn't appear to correspond to the position of any
major central dominant galaxy). Once again we applied the three
statistical tests to the two distributions to find out whether these
apparent differences are significant. Three ranges were tested: a large
range (-50$'$\,$<$\,$\Delta$RA\,$<$\,50$'$,
-50$'$\,$<$\,$\Delta$Dec\,$<$\,50$'$), a smaller central range
(-20$'$\,$<$\,$\Delta$RA\,$<$\,20$'$,
-20$'$\,$<$\,$\Delta$Dec\,$<$\,20$'$), and a range surrounding NGC 4839
where CD96 found the most significant subclustering
(-50$'$\,$<$\,$\Delta$RA\,$<$\,-10$'$,
-50$'$\,$<$\,$\Delta$Dec\,$<$\,-10$'$). The two binned tests (the
Monte-Carlo and the ${\chi}^2$) used a 7$\times$7 grid and a cutoff of
10 galaxies per bin for the large region, a 5$\times$5 grid and a cutoff
of 8 galaxies per bin for the central region, and a 4$\times$4 grid and
cutoff of 5 galaxies per bin for the region around NGC 4839. These
values were chosen based on the size of each region and the density of
galaxies there. The results of the statistical tests (once again in
terms of the percentage chance of difference) are given in Table 3. None
of the significance levels produced by any of the tests is high enough
to be held up as strong evidence for a statistically significant
difference. It could be argued that the grid size used in the binned
tests was too large to pick up the features described by B96 (namely the
difference in position of the peaks in the giant density and the dwarf
density). To check this, we have repeated the tests using twice as many
bins in the central region and cutoff of 3 galaxies per bin, and the
results are approximately the same as before (no greater than a 90\%
chance of the two distributions being different). The same is true if we
restrict attention to an even smaller central region of
(-5$'<\Delta$RA\,$<$\,10$'$, -5$<\Delta$Dec\,$<$\,10$'$) which just
encloses the density peaks for both giants and dwarfs.

\section{Discussion}

B96 concluded that the dwarf distribution was significantly more uniform
than the giant distribution. They offered as an explanation the
possibility that the original subclusters that make up Coma had time to
proceed towards equipartition before they merged, spreading out the
dwarfs in both position and velocity space. Our results for the
line-of-sight velocity distributions of the giants and dwarfs are
consistent with the findings of B96, in that there is significant
substructure (and thus non-Gaussianity) present in the giant
distribution but not the dwarf distribution. However, we find no
evidence for differences between the localised velocity distributions or
the spatial distributions of the giants and dwarfs, which contradicts
B96. The primary reason why our conclusions differ is that B96 made no
direct statistical comparisons between the giant and dwarf
distributions. Although to the eye there do appear to be differences,
our tests show that there is still a significant probability of the
giants and dwarfs belonging to the same distribution. Nonetheless, it is
worth investigating whether it is plausible that the original
subclusters could have been in equipartition, as B96 proposed.

The main processes by which energy can be transferred from massive
galaxies to lighter ones in clusters are two-body relaxation and
dynamical friction. The typical timescale for two-body relaxation in a
cluster the size of Coma is $t_{tb} \gtrsim 3 \times 10^{11}$ yr
\citep{Sar86}, much greater than a Hubble time, so we can expect that
dynamical friction will be the dominant process driving the cluster
towards equipartition. If we assume that the core of the cluster is an
isothermal sphere, the timescale $t_{fr}$ for dynamical friction to
significantly alter the momentum of a galaxy with mass M is given by
\citep{Bin87}:
\begin{equation}
t_{fr} = \frac{66 \ \textrm{Gyrs}}{\ln \Lambda} \left( \frac{r_i}{0.5 \ \textrm{Mpc}}
\right) ^2 \left( \frac{\sigma}{1000 \ \textrm{km s$^{-1}$}} \right) \left( 
\frac{10^{12} \  \textrm{M}_{\odot}}{\textrm{M}} \right)
\label{df}
\end{equation}
where $\Lambda$ is the ratio of the largest possible impact parameter to
the smallest, $r_i$ is the initial radius of the galaxy's orbit, and
$\sigma$ is the velocity dispersion of the cluster. Typically $\ln
\Lambda \sim 3$ \citep{Sar86}; for Coma, $r_i~\simeq$~0.5 Mpc, and
$\sigma \simeq$ 1000~km~s$^{-1}$. When these values are substituted in
to Eq. \ref{df}, we find that $t_{fr} \simeq 22$ Gyr for a galaxy of
10$^{12} \ \textrm{M}_{\odot}$ (a typical mass for a `giant' galaxy). A
galaxy would have to be more massive than $2.2 \times 10^{13} \
\textrm{M}_{\odot}$ to have a dynamic friction timescale less than 1
Gyr. According to \citet{Vik94}, the masses of the two largest galaxies
in the cluster, NGC 4874 and NGC 4889, are both approximately $1 \times
10^{13} \ \textrm{M}_{\odot}$, so not even they should have had time to
settle into the centre of the cluster potential. This is consistent with
the observation that the galaxy velocities do not show the signature of
equipartition, and that the velocities of the main cD galaxies are
displaced from the mean velocity of the surrounding galaxies.

Is it plausible that the subclusters conjectured to be merging with Coma
would have had time to approach equipartition? A rough criterion for
judging this is to require $t_{fr}$ to be less than 1 Gyr for at least
the most massive galaxies in a subcluster. The velocity dispersion of a
cluster scales roughly as the square-root of its mass, so a subcluster
with around 0.1 times the mass of Coma should have a velocity dispersion
of $\sqrt{0.1} \ \times$ the dispersion of Coma, or about 300 km
s$^{-1}$. Then $r_i$ for such a subcluster only needs to be a factor of
3 smaller than that of Coma to to bring $t_{fr}$ below 1 Gyr for a
$10^{12} \ \textrm{M}_{\odot}$ galaxy. These values of $\sigma$ and
$r_i$ are well within the range that have actually been observed in
small clusters \citep{Ada98}, so there is a good chance that any
subclusters that have merged with Coma would have been at least
partially in equipartition.

The next question in this: if Coma is really composed of subclusters
that are (or were) in equipartition, are our statistical tests capable
of detecting it? We checked this by carrying out some Monte-Carlo
simulations. The first set, described above in the section 3.1,
reassigned velocities to each of the galaxies in a way consistent with
the \emph{entire cluster} being in equipartition. The positions of the
galaxies were left unchanged. The resulting velocity distributions
showed a marked difference between giants and dwarfs, much larger than
the difference seen in the real data. We also tested the
D-vs-\textit{cz} distributions from the simulated data, in exactly the
same way as was done for the real data. The result (for the large range)
was always a $>$99.9\% probability of the giant and dwarf distributions
being different (the results in the central range were a bit less
consistent, with $\sim$ 1/2 of the simulations resulting in a $>$99\%
significance level). So we can be fairly confident that if the entire
cluster was in equipartition we would be able to detect it. The fact
that we don't is consistent with our expectation that Coma as a whole is
too large to have approached equipartition.

The second set of simulations were done to mimic two equipartitioned
sub-clusters (centred around NGC 4889 and NGC 4839 respectively) in the
process of merging with a `main cluster body' centred on NGC 4874. This
is approximately the merger scenario proposed by CD96. A certain
fraction of the galaxies within 25$'$ of NGC 4889 and NGC 4839 were
randomly assigned as being `bound'; the fraction was chosen so that the
number of galaxies bound to each of NGC 4889 and NGC 4839 comprised
10\% of the total cluster population. The galaxies of these two bound
populations were assigned velocities consistent with equipartition, with
a dispersion scaled to 300~km~s$^{-1}$ and a mean velocity equal to that
of the relevant cD galaxy. We use 300~km~s$^{-1}$ for the dispersion of
the bound groups because these groups have about 10\% of the mass of
the entire cluster, and we expect the dispersion to scale as the square
root of the mass; $\sqrt{0.1} \simeq 0.32$, so the bound group
dispersion should be roughly one-third that of the entire cluster. The
remaining 80\% of the cluster population was assumed \emph{not} to be
in equipartition (due to the combined effects of earlier mergers,
violent relaxation and infall); these galaxies were assigned velocities
taken at random from a single Gaussian with a dispersion of
1000~km~s$^{-1}$ and a mean equal to the velocity of NGC 4874.

One thousand of these `merger' velocity distributions were simulated.
For each, the line-of-sight velocity distributions for giants and dwarfs
had very similar dispersions ($\sim 1000$~km~s$^{-1}$), just like in the
real data. Direct comparisons between the giant and dwarf velocity
distributions using a K-S test failed to turn up a significant
difference. Remarkably, though, for 37\% of the simulated distributions
the velocities of the giants were non-Gaussian with a $>$99\%
certainty. Conversely, only 1.4\% of the dwarf distributions were
significantly non-Gaussian. The reason for this is that the effect of
equipartition in the bound groups is to reduce the dispersion of giant
velocities and spread out the dwarf velocities, so there tend to be
distinct peaks in the giant line-of-sight velocity distribution around
the velocities of NGC 4889 and NGC 4839 but not in the dwarf
distribution. Thus it is plausible that this sort of merger scenario
could account for the non-Gaussianity of the real giant distribution and
the contrasting Gaussianity of the real dwarf distribution.

The D-vs-\textit{cz} distributions of the simulated data were also
compared, using the 2D K-S two-sample test. The giant and dwarf
distributions were found to have on average an 85\% chance of being
different; only 2.3\% of the simulated distributions were found to have
a $>$99\% chance of being different. Thus the statistical tests fail to
detect the signature of equipartition in the subclusters with a high
level of confidence, just as they did with the real data.

We would expect that, in principle at least, increasing the number of
redshifts in our sample would boost our chances of detecting evidence of
differences between the giants and dwarfs. However, there are only so
many galaxies there to be measured. We estimate that roughly 670 of the
dwarf galaxies listed in the GMP catalogue, which is complete to b\,=\,20,
are actually cluster members. The reasoning behind this is that about
52\% of the galaxies with measured redshifts (and bluer than the top
edge of the cluster colour-magnitude relation) are actually cluster
members, so roughly the same percentage of the total number of dwarfs in
the GMP catalogue (again, bluer than the top edge of the cluster
colour-magnitude relation) are likely to be cluster members. If we had
redshifts for all 670 of these dwarfs, would we then be able to detect
significant differences between the giants and the dwarfs? To answer
this question, we ran a final set of merger-mimicking simulations, this
time including all the giants and 670 dwarfs, drawn at random from the
catalogue. The chances of detecting differences did improve slightly
with the increased sample size, but still not enough to be statistically
significant. The 2D K-S two-sample test found on average that there was
a 87\% chance that the giant and dwarf D-vs-\textit{cz} distributions
were different; only 5\% of the simulated distributions were found to
have a $>$99\% chance of being different. This suggests that even if we
measured redshifts for every dwarf in the cluster brighter than b\,=\,20,
we would still not have a high chance of detecting evidence of
equipartitioned subclusters in the D-vs-\textit{cz} distributions.

Even though this toy model of a merger scenario is contrived, it
demonstrates that Coma \emph{could} indeed consist of equipartitioned
subclusters. Indeed, the subcluster equipartitioning seems to generate
the sort of non-Gaussianity in the giants (but not the dwarfs) that we
see in the real data. However, it doesn't generate differences between
the giant and dwarf distributions that are strong enough to detect with
our direct statistical comparisons, even if we measured the redshifts of
all dwarf cluster members down to b\,=\,20.
 
\section{Conclusion}

The distribution of line-of-sight velocities for giant galaxies (those
with b\,$<$\,18) in the Coma cluster shows strong evidence of
non-Gaussianity, caused primarily by a large gap in velocities around
\textit{cz} = 6500~km~s$^{-1}$. This gap, and others in the
distribution, seem to correspond to the velocities of the cD galaxies
NGC 4489 and NGC 4839, but it is still a mystery whether or not the cD
galaxies are directly responsible for causing the gaps. Unlike that for
the giants, the distribution of velocities for a sample of dwarf
galaxies (b\,$>$\,18) exhibits no significant non-Gaussianity and no
significant gaps. However, tests directly comparing the localised
velocity substructure and spatial substructure in the giant and dwarf
distributions show no strong evidence for significant differences
between the two populations. Our results rule out the cluster as a whole
having moved significantly towards equipartition. On the other hand,
they are consistent with the possibility that Coma formed from the
merger of subclusters that were in equipartition; indeed, our
simulations suggest that such a scenario can explain the non-Gaussianity
of the giant line-of-sight velocity distribution and the contrasting
Gaussianity of the dwarfs. We also find that such a merger scenario may
not generate a strong enough difference between the giant and dwarf
distributions to be able to detect by direct statistical comparison,
even if the redshifts of all cluster members brighter than b\,=\,20 were
measured, so more sensitive tests must be developed to put stronger
constraints on Coma's merger history.

\clearpage
\begin{center}
\begin{tabular}{l c c c}
\multicolumn{4}{c}{\textsc{TABLE 1}} \\
\multicolumn{4}{c}{\textsc{Results for D-vs-\textit{cz}}} \\
\tableline \tableline
Range         & ${\chi}^2$ & K-S & Monte-Carlo \\
\tableline
Large range   & 81\% & 71\% & 81\% \\    
Central range & 15\% & 85\% & 28\% \\
\tableline
\end{tabular}
\end{center}
\vspace{4ex}
\begin{center}
\begin{tabular}{l c c c}
\multicolumn{4}{c}{\textsc{TABLE 2}} \\
\multicolumn{4}{c}{\textsc{Results for velocity slices}} \\
\tableline \tableline
Range         & ${\chi}^2$ & K-S & Monte-Carlo \\
\tableline
A (5400$<$\textit{cz}$<$6600 km s$^{-1}$)  & 64\% & 17\% & 36\% \\    
B (6400$<$\textit{cz}$<$7600 km s$^{-1}$)  & 22\% & 19\% & 10\% \\
C (7400$<$\textit{cz}$<$8600 km s$^{-1}$)  & 27\% & 84\% & 24\% \\
\tableline
\end{tabular}
\end{center}
\vspace{4ex}
\begin{center}
\begin{tabular}{l c c c}
\multicolumn{4}{c}{\textsc{TABLE 3}} \\
\multicolumn{4}{c}{\textsc{Results for projected position on the sky}} \\
\tableline \tableline
Range           & ${\chi}^2$ & K-S & Monte-Carlo \\
\tableline
Large range     & 71\% & 92\% & 83\% \\    
Central range   & 59\% & 35\% & 86\% \\
Around NGC 4839 & 57\% & 55\% & 34\% \\
\tableline
\end{tabular}
\end{center}



\begin{figure}
\begin{center}
\plotone{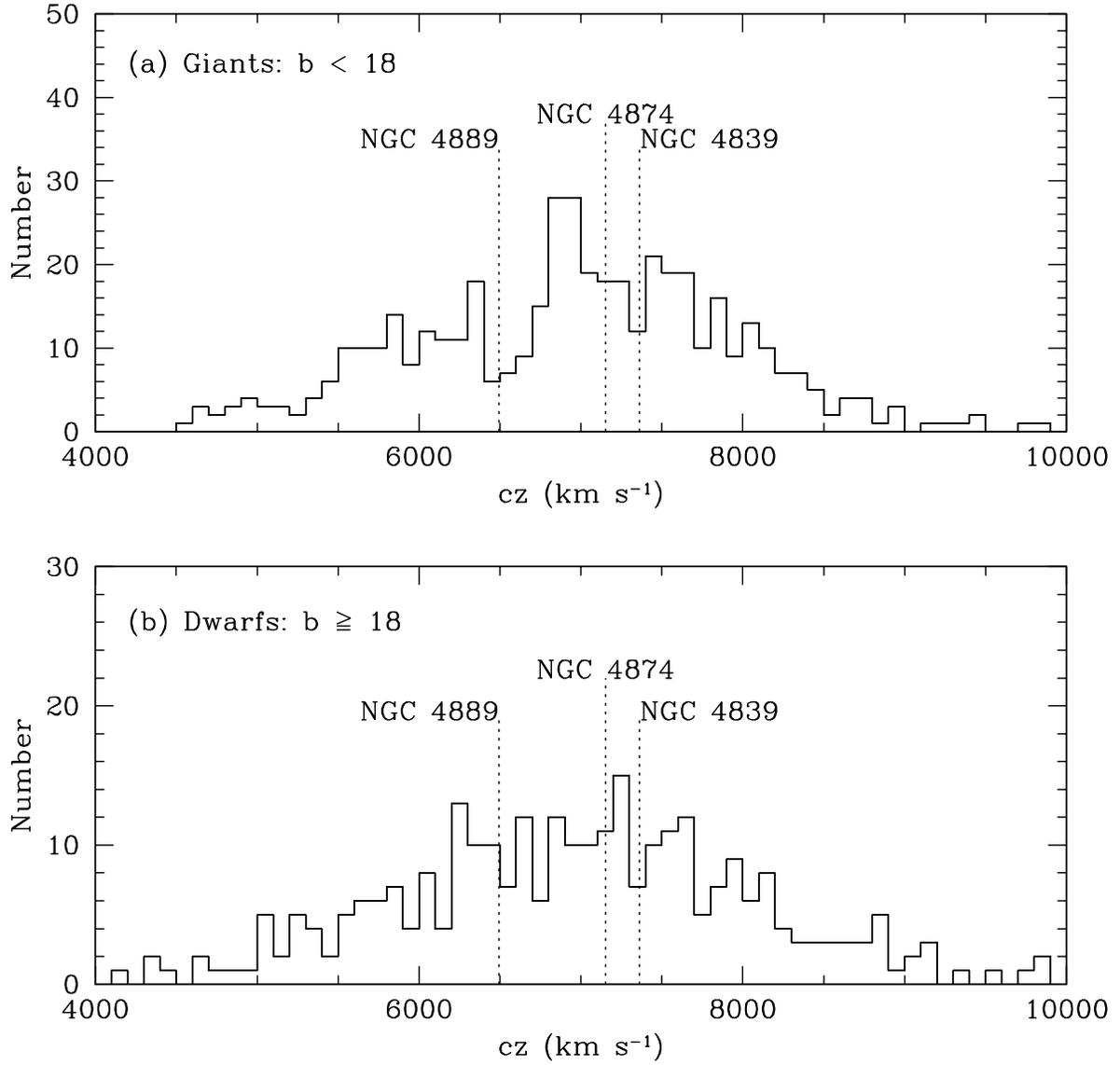}
\caption{Histograms of the line-of-of sight velocity distribution for
galaxies in the Coma cluster. (a) Giants (b\,$<$\,18). (b) Dwarfs
(b\,$>$\,18).}
\end{center}
\label{hist}
\end{figure}


\begin{figure}
\begin{center}
\plotone{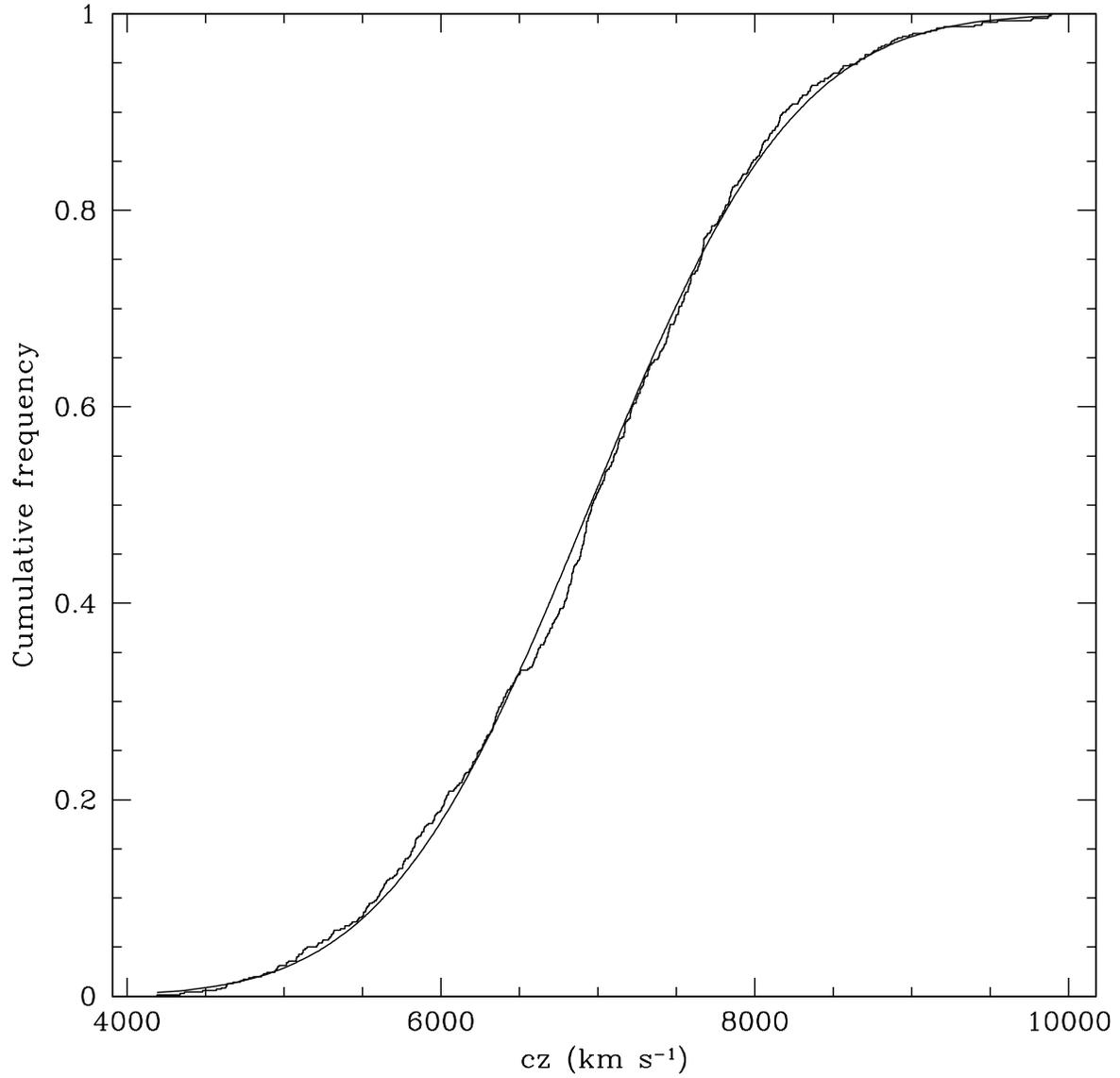}
\caption{Cumulative frequency of giants as a function of redshift
(stepped line) superimposed on a cumulative Gaussian (smooth curve) that
has the same mean and dispersion as the giants. Note that the largest
gap between the two, which defines the K-S statistic, is at \textit{cz}
$\sim$ 6500~km~s$^{-1}$.}
\end{center}
\label{cumugauss}
\end{figure}


\begin{figure}
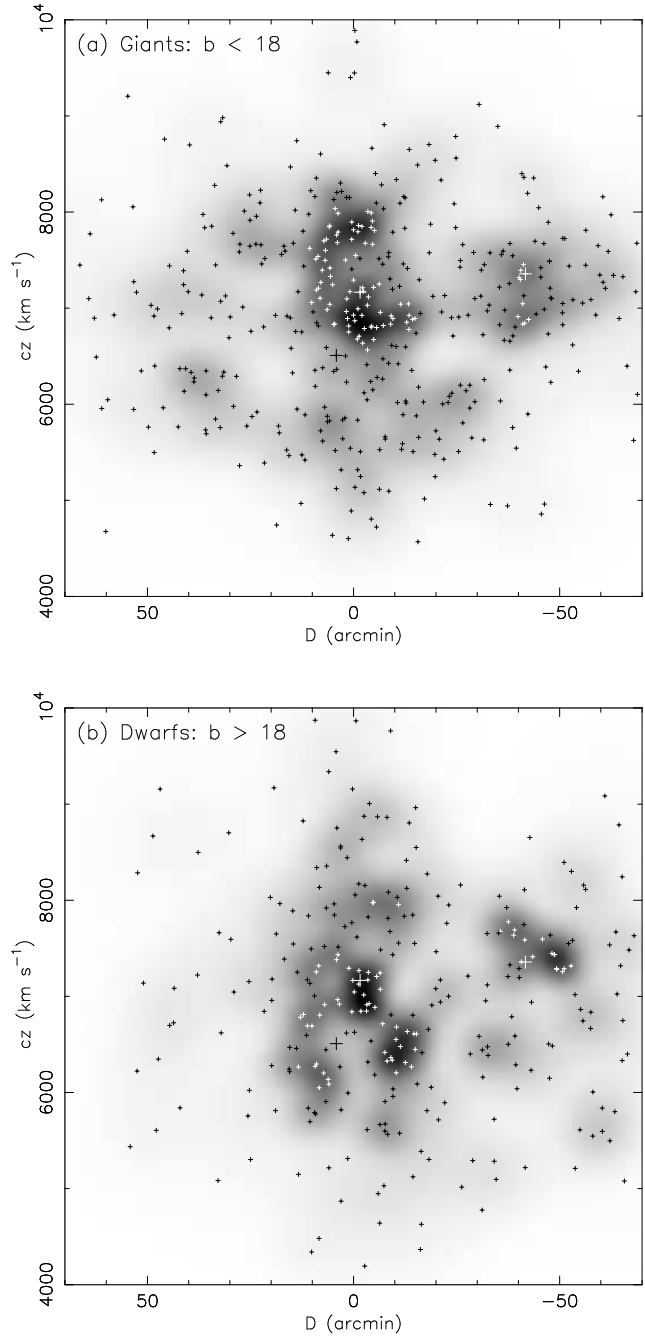

\begin{center}
\epsscale{0.51}
\plotone{f3_a.eps}
\makebox[\textwidth][c]{}
\plotone{f3_b.eps}
\caption{Smoothed density of galaxies as a function of line-of-sight
velocity \textit{cz} and distance D along the NE-SW axis of the cluster
(where NE is positive). (a) is the distribution of giants, and (b) is
the distribution of dwarfs. In each case the density is smoothed using
an adaptive 2D Gaussian kernel with base dispersions ${\sigma}_D$ = 5$'$
and ${\sigma}_{\textit{cz}}$ = 200~km~s$^{-1}$. The small crosses mark
the positions of the galaxies; the three large crosses represent the
three dominant galaxies: from left to right, NGC 4889, NGC 4874 and NGC
4839.}
\end{center}
\label{Dcz}
\end{figure}


\begin{figure}
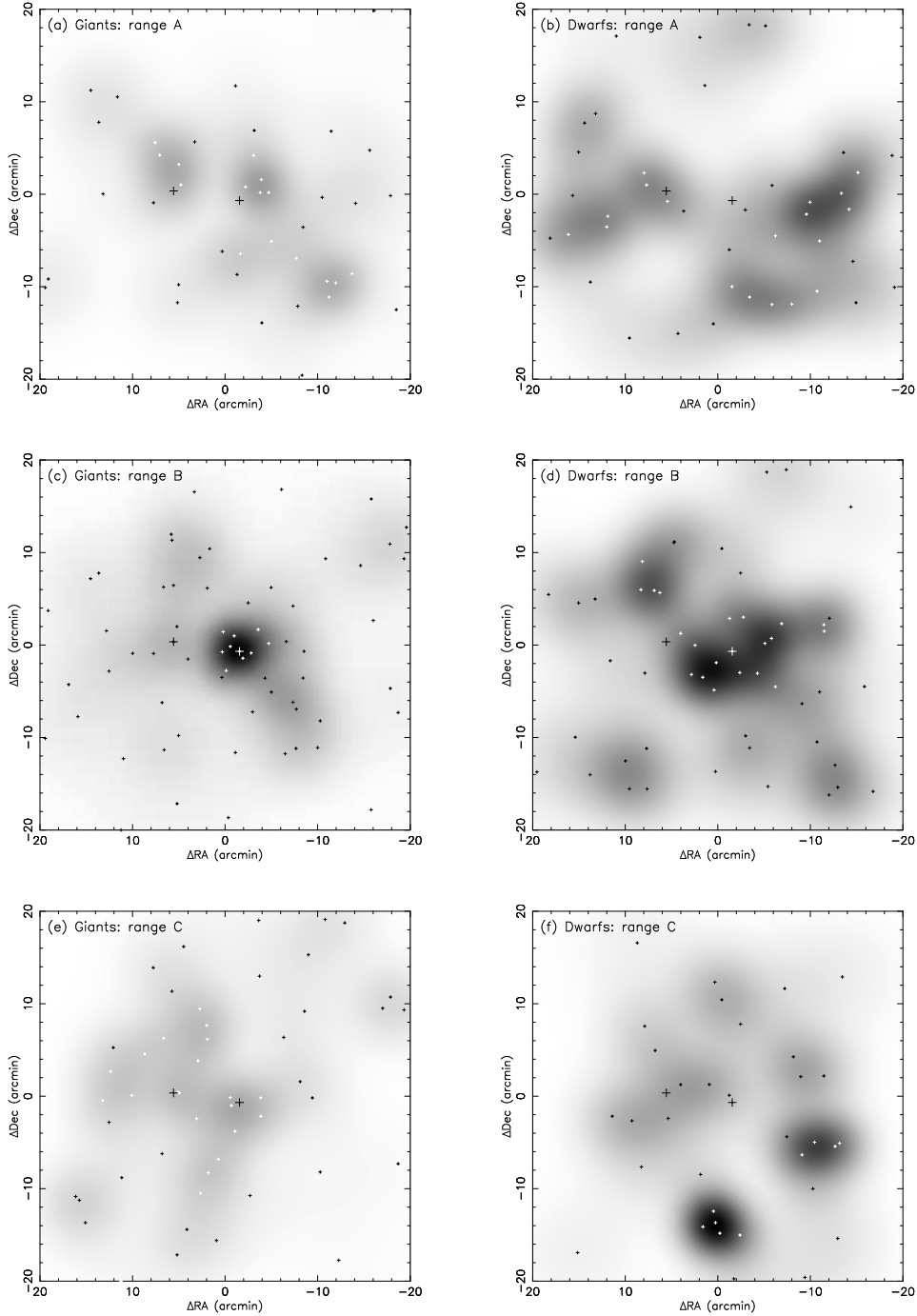

\begin{center}
\epsscale{0.35}
\plotone{f4_a.eps}
\makebox[0.05\textwidth][c]{}
\plotone{f4_b.eps}
\makebox[\textwidth][s]{ }
\plotone{f4_c.eps}
\makebox[0.05\textwidth][c]{}
\plotone{f4_d.eps}
\makebox[\textwidth][s]{ }
\plotone{f4_e.eps}
\makebox[0.05\textwidth][c]{}
\plotone{f4_f.eps}
\caption{Six slices in velocity through the
$\Delta$RA-$\Delta$Dec-\textit{cz} distribution: (a), (c) and (e) are
the distribution of giants in the velocity ranges A, B and C
respectively; (b), (d) and (f) are the same but for the dwarfs. The
three ranges are A: (5400\,$<$\,\textit{cz}\,$<$\,6600), B:
(6400\,$<$\,\textit{cz}\,$<$\,7600), and C:
(7400\,$<$\,\textit{cz}\,$<$\,8600), where \textit{cz} is
in~km~s$^{-1}$. Each density plot was smoothed with an adaptive 2D
Gaussian kernel with base dispersions ${\sigma}_{\Delta\mathrm{RA}}$ =
${\sigma}_{\Delta\mathrm{Dec}}$ = 3$'$. The small crosses are the
positions of the galaxies, and the two large crosses are NGC 4889 (left)
and NGC 4874 (right).}
\end{center}
\label{slices}
\end{figure}


\begin{figure}
\epsscale{1}
\plottwo{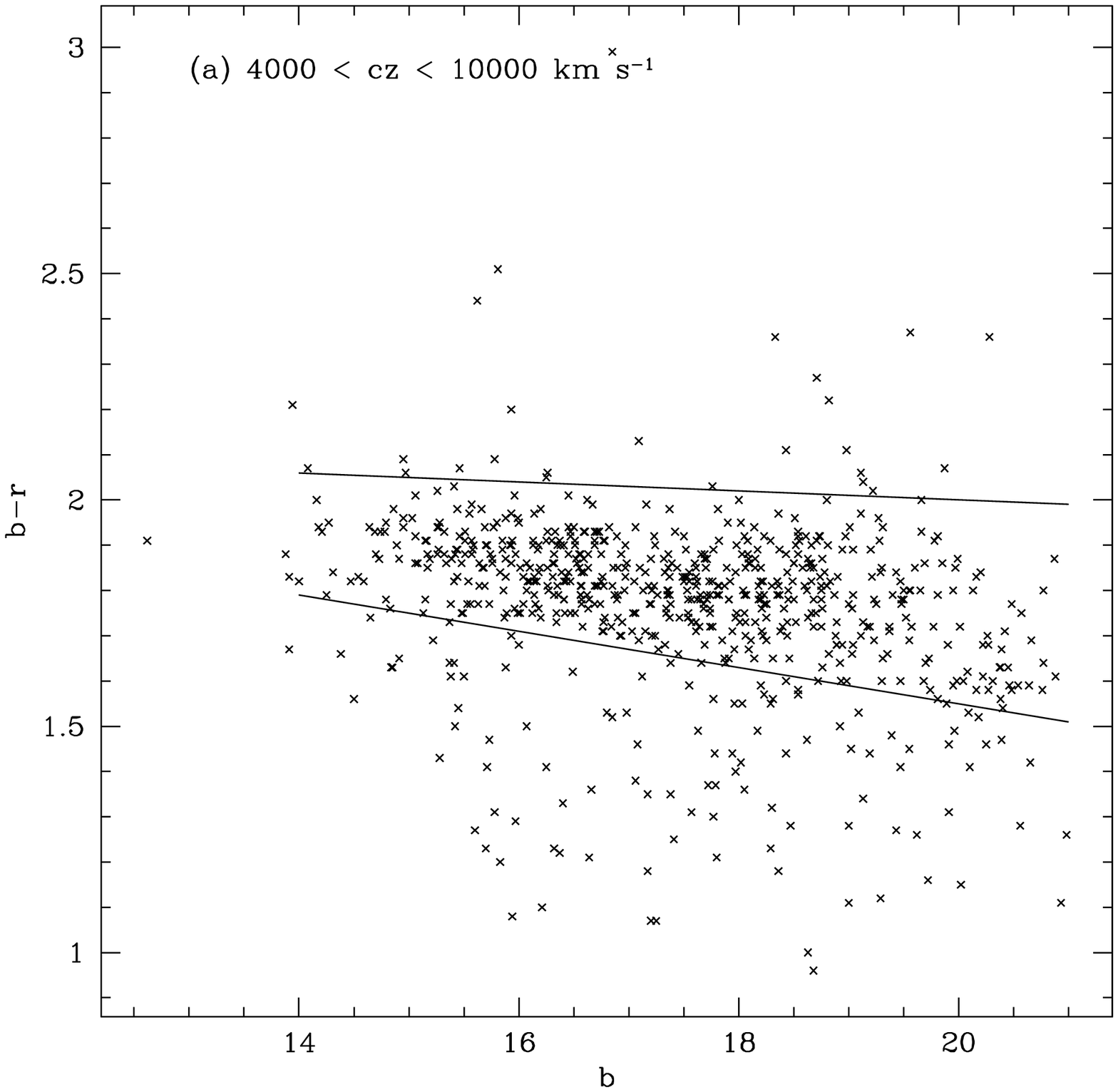}{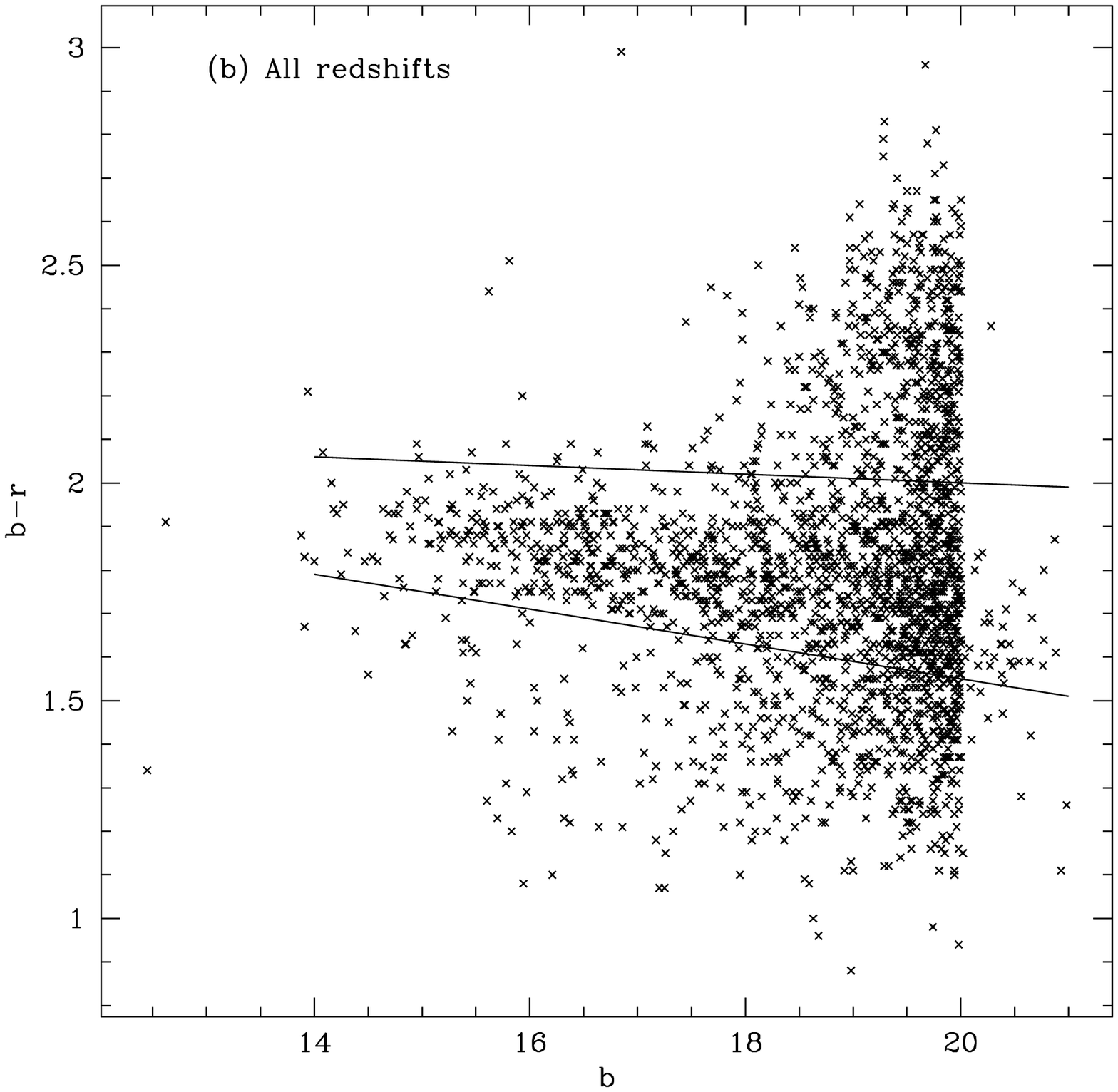}
\caption{(a) b-r colour plotted against magnitude b for all galaxies
within the redshift range
(4000\,$<$\,\textit{cz}\,$<$\,10000)~km~s$^{-1}$ (and thus deemed to be
members of the cluster). The solid lines delineate the band within which
most of the points fall. (b) The same plot for all galaxies, regardless
of redshift. Any galaxies without measured redshifts, and that fall
outside the cluster colour-magnitude relation determined from (a), are
assumed to be field galaxies and are neglected.}
\label{colmag}
\end{figure}


\begin{figure}
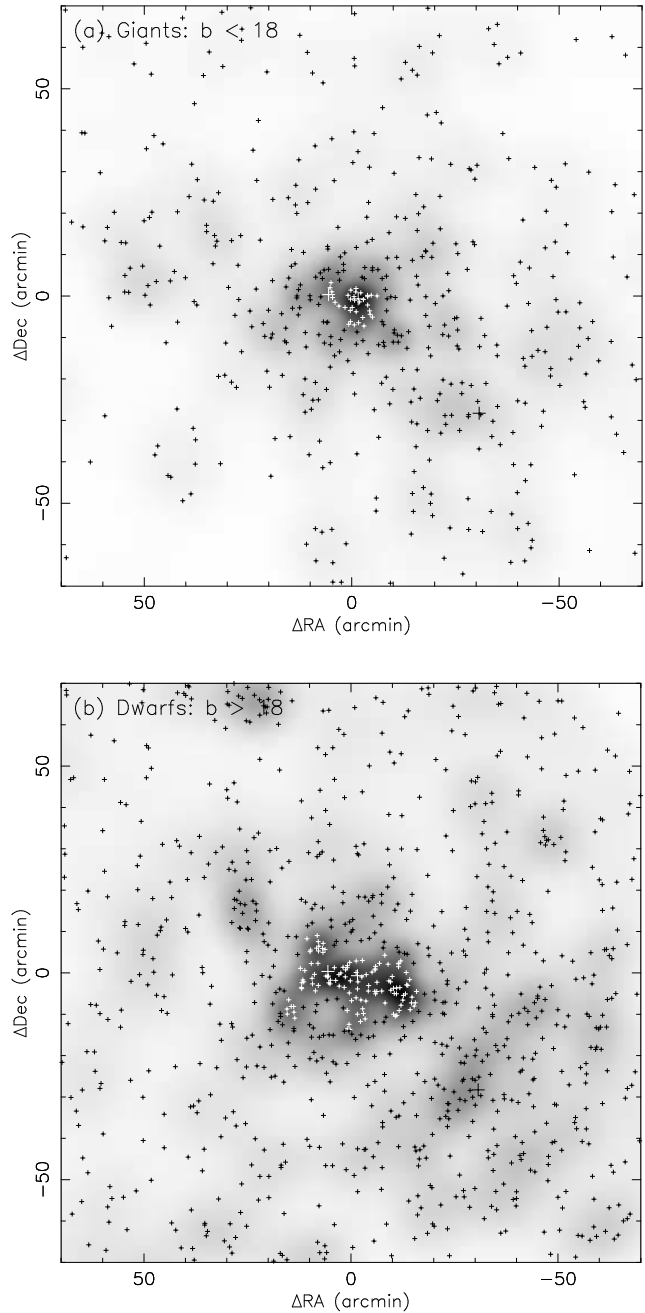

\epsscale{0.51}
\plotone{f6_a.eps}
\makebox[\textwidth][c]{}
\plotone{f6_b.eps}
\caption{Density plots of spatial position for giants (a) and dwarfs
(b). The galaxy samples used here include all galaxies with
4000\,$<$\,\textit{cz}\,$<$\,10000~km~s$^{-1}$ plus galaxies without
redshifts but within the cluster colour-magnitude relation (see Fig. 5).
Smoothing is done with an adaptive 2D Gaussian kernel with base
dispersions ${\sigma}_{\Delta\mathrm{RA}}$ =
${\sigma}_{\Delta\mathrm{Dec}}$ = 5$'$. As before, the small crosses
mark the positions of the galaxies and the large crosses mark NGC 4889,
NGC 4874 and NGC 4839 (from left to right).}
\label{spatialtot}
\end{figure}

\end{document}